\documentclass[aps,showpacs,twocolumn,amsmath,amssymb,superscriptaddress,prb]{revtex4-1}

\usepackage{graphicx}
\usepackage{bm}
\usepackage{color} 

\begin{document}

\title{Spin absorption at ferromagnetic-metal/platinum-oxide interface}

\author{Akio Asami} \affiliation{Department of Applied Physics and Physico-Informatics, Keio University, Yokohama 223-8522, Japan}

\author{Hongyu An}
\affiliation{Department of Applied Physics and Physico-Informatics, Keio University, Yokohama 223-8522, Japan}

\author{Akira Musha}
\affiliation{Department of Applied Physics and Physico-Informatics, Keio University, Yokohama 223-8522, Japan}

\author{Makoto Kuroda}
\affiliation{Department of Applied Physics and Physico-Informatics, Keio University, Yokohama 223-8522, Japan}

\author{Kazuya Ando}
\email{ando@appi.keio.ac.jp}
\affiliation{Department of Applied Physics and Physico-Informatics, Keio University, Yokohama 223-8522, Japan}

\date{\today}

\begin{abstract}
We investigate the absorption of a spin current at a ferromagnetic-metal/Pt-oxide interface by measuring current-induced ferromagnetic resonance. The spin absorption was characterized by the magnetic damping of the heterostructure. We show that the magnetic damping of a Ni$_{81}$Fe$_{19}$ film is clearly enhanced by attaching Pt-oxide on the Ni$_{81}$Fe$_{19}$ film. The damping enhancement is disappeared by inserting an ultrathin Cu layer between the Ni$_{81}$Fe$_{19}$ and Pt-oxide layers. These results demonstrate an essential role of the direct contact between the Ni$_{81}$Fe$_{19}$ and Pt-oxide to induce sizable interface spin-orbit coupling. Furthermore, the spin-absorption parameter of the Ni$_{81}$Fe$_{19}$/Pt-oxide interface is comparable to that of intensively studied heterostructures with strong spin-orbit coupling, such as an oxide interface, topological insulators, metallic junctions with Rashba spin-orbit coupling. This result illustrates strong spin-orbit coupling at the ferromagnetic-metal/Pt-oxide interface, providing an important piece of information for quantitative understanding the spin absorption and spin-charge conversion at the ferromagnetic-metal/metallic-oxide interface. 
\end{abstract}

\maketitle

\section{introduction}

An emerging direction in spintronics aims at discovering novel phenomena and functionalities originating from spin-orbit coupling (SOC)~\cite{soumyanarayanan2016emergent}. An important aspect of the SOC is the ability to convert between charge and spin currents. The charge-spin conversion results in the generation of spin-orbit torques in heterostructures with a ferromagnetic layer, enabling manipulation of magnetization~\cite{gambardella2011current,qiu2018characterization,brataas2012current}. Recent studies have revealed that the oxidation of the heterostructure strongly influences the generation of the spin-orbit torques. The oxidation of the ferromagnetic layer alters the spin-orbit torques, which cannot be attributed to the bulk spin Hall mechanism~\cite{qiu2015spin,hibino2017enhancement,PhysRevLett.121.017202}. The oxidation of a nonmagnetic layer in the heterostructure also offers a route to engineer the spin-orbit devices. Demasius $et$ $al$. reported a significant enhancement of the spin-torque generation by incorporating oxygen into tungsten, which is attributed to the interfacial effect~\cite{demasius2016enhanced}. The spin-torque generation efficiency was found to be significantly enhanced by manipulating the oxidation of Cu, enabling to turn the light metal into an efficient spin-torque generator, comparable to Pt~\cite{an2016spin}. We also reported that the oxidation of Pt turns the heavy metal into an electrically insulating generator of the spin-orbit torques, which enables the electrical switching of perpendicular magnetization in a ferrimagnet sandwiched by insulating oxides~\cite{an2018current}. These studies have provided valuable insights into the oxide spin-orbitronics and shown a promising way to develop energy-efficient spintronics devices based on metal oxides.

The SOC in solids is responsible for the relaxation of spins, as well as the conversion between charge and spin currents. The spin relaxation due to the bulk SOC of metals and semiconductors has been studied both experimentally and theoretically~\cite{mizukami2002effect,tserkovnyak2002spin,chen2013direct,nakayama2012geometry}. The influence of the SOC at interfaces on spin-dependent transport has also been recognized in the study of giant magnetoresistance (GMR). The GMR in Cu/Pt multilayers in the current-perpendicular-to-plane  geometry indicated that there must be a significant spin-memory loss due to the SOC at the Cu/Pt interfaces~\cite{kurt2002spin}. The interface SOC also plays a crucial role in recent experiments on spin pumping. The spin pumping refers to the phenomenon in which precessing magnetization emits a spin current to the surrounding nonmagnetic layers~\cite{tserkovnyak2002spin}. When the pumped spin current is absorbed in the nonmagnetic layer due to the bulk SOC or the ferromagnetic/nonmagnetic interface due to the interface SOC, the magnetization damping of the ferromagnetic layer is enhanced because the spin-current absorption deprives the magnetization of the angular momentum~\cite{tserkovnyak2002enhanced}. Although the damping enhancement induced by the spin pumping has been mainly associated with the spin absorption in the bulk of the nonmagnetic layer, recent experimental and theoretical studies have demonstrated that the spin-current absorption at interfaces also provides a dominant contribution to the damping enhancement~\cite{sanchez2013spin}. Since the absorption of a spin current at interfaces originates from the SOC, quantifying the damping enhancement provides an important information for fundamental understanding of the spin-orbit physics.

In this work, we investigate the absorption of a spin current at a ferromagnetic-metal/Pt-oxide interface. We show that the magnetic damping of a Ni$_{81}$Fe$_{19}$ (Py) film is clearly enhanced by attaching Pt-oxide, Pt(O), despite the absence of the absorption of the spin current in the bulk of the Pt(O) layer. The damping enhancement disappears by inserting an ultrathin Cu layer between the Py and Pt(O) layers. This result indicates that the direct contact between the ferromagnetic metal and Pt oxide is essential to induce the sizable spin-current absorption, or the interface SOC. We further show that the strength of the damping enhancement observed for the Py/Pt(O) bilayer is comparable with that reported for other systems with strong SOC, such as two-dimensional electron gas (2DEG) at an oxide interface and topological insulators.

\section{experimental methods}
Three sets of samples, Au/SiO$_2$/Py, Au/SiO$_2$/Py/Pt(O) and Au/SiO$_2$/Py/Cu/Pt(O), were deposited on thermally oxidized Si substrates (SiO$_2$) by RF magnetron sputtering at room temperature. To avoid the oxidation of the Py or Cu layer, we first deposited the Pt(O) layer on the SiO$_2$ substrate in a mixed argon and oxygen atmosphere. After the Pt(O) deposition, the chamber was evacuated to 1$\times10^{-6}$ Pa, and then the Py or Cu layer was deposited on the top of the Pt(O) layer in a pure argon atmosphere. For the Pt(O) deposition, the amount of oxygen gas in the mixture was fixed as 30$\%$, in which the flow rates of argon and oxygen were set as 7.0 and 3.0 standard cubic centimeters per minute (sccm), respectively. The SiO$_2$ layer was deposited from a SiO$_2$ target in the pure argon atmosphere. The film thickness was controlled by the deposition time with a precalibrated deposition rate.

We measured the magnetic damping using current-induced ferromagnetic resonance (FMR). For the fabrication of the devices used in the FMR experiment, the photolithography and lift-off technique were used to pattern the films into a 10 $\mu$m $\times$ 40 $\mu$m rectangular shape. A blanket Pt(O) film on a 1 cm $\times$ 1 cm SiO$_2$ substrate was fabricated for the composition confirmation by x-ray photoelectron spectroscopy (XPS). We also fabricated Pt(O) single layer and SiO$_2$/Py/Pt(O) multilayer films with a Hall bar shape to determine the resistivity of the Pt(O) and Py using the four-probe method. The resistivity of Pt(O) (6.3 $\times10^6$ $\mu\Omega$ cm) is much larger than that of Py (106 $\mu\Omega$ cm). Because of the semi-insulating nature of the Pt(O) layer, we neglect the injection of a spin current into the Pt(O) layer from the Py layer; only the Py/Pt(O) interface can absorb a spin current emitted from the Py layer. Transmission electron microscopy (TEM) was used to directly observe the interface and multilayer structure of the SiO$_2$/Py/Pt(O) film.  All the measurements were conducted at room temperature.

\section{results and discussion}
Figure 1(a) exhibits the XPS spectrum of the Pt(O) film. Previous XPS studies on Pt(O) show that binding energies of the Pt 4$f_{7/2}$ peak for Pt, PtO and PtO$_2$ are around 71.3, 72.3 and 74.0 eV, respectively~\cite{wang2014polyol}. Thus, the Pt 4$f_{7/2}$ peak at 72.3 eV in our Pt(O) film indicates the formation of PtO. By further fitting the XPS spectra, we confirm that the Pt(O) film is composed of a dominant structure of PtO with a minor portion of PtO$_2$.             
Figure 1(b) shows the cross-sectional TEM image of the SiO$_2$(4 nm)/Py(8 nm)/Pt(O)(10 nm) film. As can be seen, continuous layer morphology with smooth and distinct interfaces is formed in the multilayer film. Although we deposited the Py layer on the Pt(O) layer to avoid the oxidation of the Py, it might still be possible that the Py layer is oxidized by the Pt(O) layer. Therefore, we measured the resistance of the Au/SiO$_2$/Py and Au/SiO$_2$/Py/Pt(O) samples used in the FMR experiment. The resistance of both samples show the same value (60 $\Omega$). Furthermore, as described in the following section, the saturation magnetization for each device was obtained by using Kittel formula (0.746 T and 0.753 T for the Au/SiO$_2$/Py and Au/SiO$_2$/Py/Pt(O), respectively). The only 1$\%$ difference indicates that the minor oxidation of the Py layer due to the presence of the Pt(O) layer can be neglected.

\begin{figure}[tb]
\includegraphics[scale=1]{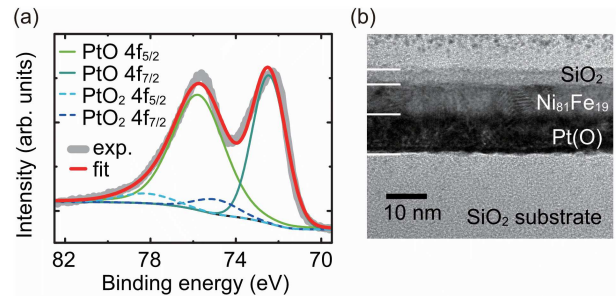}
\caption{(a) The XPS spectrum of the Pt(O) film. The gray curve is the experimental data, and the red fitting curve is the merged PtO and PtO$_{2}$ peaks. (b) The cross-sectional TEM image of the SiO$_2$(4 nm)/Ni$_{81}$Fe$_{19}$(8 nm)/Pt(O)(10 nm) film.
}
\label{fig1} 
\end{figure}

\begin{figure}[tb]
\includegraphics[scale=1]{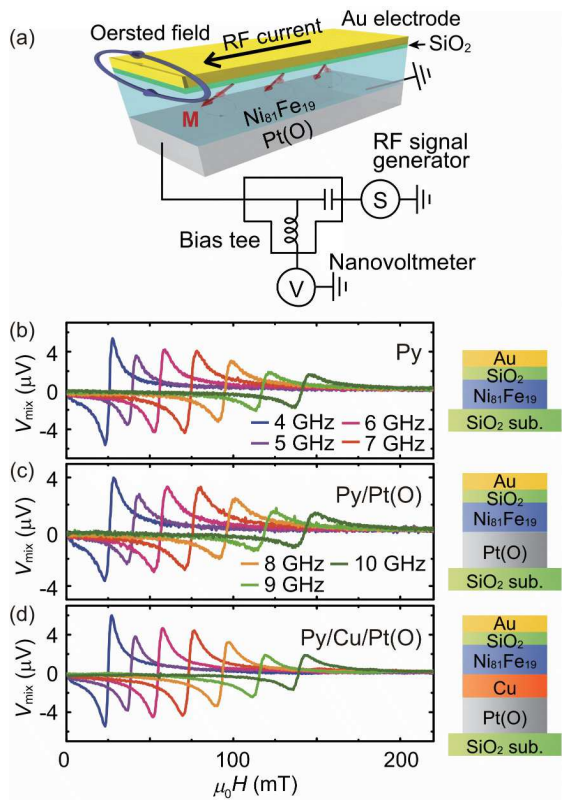}
\caption{ 
Schematic illustration of the experimental setup for the current-induced FMR. ${\bf M}$ is the magnetization in the Py layer. The FMR spectra of the (b) Py(9 nm), (c) Py(9 nm)/Pt(O)(7.3 nm), and (d) Py(9 nm)/Cu(3.6 nm)/Pt(O)(7.3 nm) films by changing the RF current frequency from 4 to 10 GHz. All the films are capped with 3 nm-thick SiO$_2$ and 10 nm-thick Au layers. The RF current power was set as 5 mW. The schematic illustrations of the corresponding films are also shown. 
}
\label{fig2} 
\end{figure}

Next, we conduct the FMR experiment to investigate the absorption and relaxation of spin currents induced by the spin pumping. Figure 2(a) shows a schematic of the experimental setup for the current-induced FMR. We applied an RF current to the device, and an in-plane external magnetic field $\mu_0H$ was swept with an angle of 45$^\text{o}$ from the longitudinal direction. The RF charge current flowing in the Au layer generates an Oersted field, which drives magnetization precession in the Py layer at the FMR condition. The magnetization precession induces an oscillation of the resistance of the device due to the anisotropic magnetoresistance (AMR) of the Py layer. We measured DC voltage generated by the mixing of the RF current and the oscillating resistance using a bias tee.

Figures 2(b), 2(c) and 2(d) show the FMR spectra for the Au/SiO$_2$/Py, Au/SiO$_2$/Py/Pt(O) and Au/SiO$_2$/Py/Cu/Pt(O) films, respectively. For the FMR measurement, a small RF current power $P=5$ mW was applied. Around $P=5$ mW, the FMR linewidth is independent of the RF power as shown in the inset to Fig.~3(a). This confirms that the measured linewidth is unaffected by additional linewidth broadening due to nonlinear damping mechanisms and Joule heating. As shown in Fig. 2, clear FMR signals with low noise are obtained, allowing us to precisely fit the spectra and extract the magnetization damping for the three samples. Here, the mixing voltage due to the FMR, $V_{\mathrm{mix}}$, is expressed as
\begin{eqnarray}
\begin{split}
V_{\mathrm{mix}}	=	V_{\mathrm{sym}}\frac{(\mu_0 \Delta H)^2}{(\mu_0 H-\mu_0 H_{\mathrm{R}})^2+(\mu_0 \Delta H)^2}\\
+V_{\mathrm{asy}}\frac{\mu_0 \Delta H(\mu_0 H-\mu_0 H_{\mathrm{R}})}{(\mu_0 H-\mu_0 H_{\mathrm{R}})^2+(\mu_0 \Delta H)^2}	\label{mixV},
\end{split}
\end{eqnarray}
where $\mu_0 \Delta H$ and $\mu_0H_\text{R}$ are the spectral width and resonance field, respectively~\cite{fang2011spin}. $V_{\mathrm{sym}}$ and $V_{\mathrm{asym}}$ are the magnitudes of the symmetric and antisymmetric components. The symmetric and antisymmetric components arise from the spin-orbit torques and Oersted field. In the devices used in the present study, the Oersted field created by the top Au layer dominates the RF effective fields acting on the magnetization in the Py layer [see also Fig.~2(a)]. The large Oersted field enables the electric measurement of the FMR even in the absence of the spin-orbit torques in the Au/SiO$_2$/Py film.

The damping constant $\alpha$ of the Py layer in the Au/SiO$_2$/Py, Au/SiO$_2$/Py/Pt(O) and Au/SiO$_2$/Py/Cu/Pt(O) films can be quantified by fitting the RF current frequency $f$ dependence of the FMR spectral width $\mu_0 \Delta H$ using  
\begin{eqnarray}
\mu_0 \Delta H	=	 \mu_0 \Delta H_\mathrm{ext} +\frac{2\pi\alpha}{\gamma}f	, \label{linear}
\end{eqnarray}
where $\Delta H_\mathrm{ext}$ and $\gamma$ are the inhomogeneous linewidth broadening of the extrinsic contribution and gyromagnetic ratio, respectively~\cite{fang2011spin,heinrich1985fmr}. Figure~3(a) shows the $f$ dependence of the FMR linewidth $\mu_0 \Delta H$, determined by fitting the spectra shown in Fig.~2 using Eq.~(\ref{mixV}). As shown in Fig. 3(a), the frequency dependence of the linewidth is well fitted by Eq.~(\ref{linear}). Importantly, the slope of the $f$ dependence of $\mu_0 \Delta H$ for the Py/Pt(O) film is clearly larger than that for the Py and Py/Cu/Pt(O) films. This indicates larger magnetic damping in the Py/Pt(O) film. By using Eq.~(2), we determined the damping constant $\alpha$ as 0.0126, 0.0169 and 0.0124 for the Py, Py/Pt(O) and Py/Cu/Pt(O) films, respectively. The difference in $\alpha$ between the  Py and Py/Cu/Pt(O) films is vanishingly small, which is within an experimental error. In contrast, the damping of the Py/Pt(O) film is clearly larger than that of the other films, indicating an essential role of the Py/Pt(O) interface on the magnetization damping.

The larger magnetic damping in the Py/Pt(O) film demonstrates an important role of the direct contact between the Py and Pt(O) layers in the spin-current absorption. If the bottom layers influence the magnetic properties of the Py layer, the difference in the magnetic properties can also result in the different magnetic damping in the Au/SiO$_2$/Py, Au/SiO$_2$/Py/Pt(O) and Au/SiO$_2$/Py/Cu/Pt(O) films. However, we have confirmed that the difference in the magnetic damping is not caused by different magnetic properties of the Py layer. In Fig. 3(b), we plot the RF current frequency $f$ dependence of the resonance field $\mu_0H_\text{R}$. As can be seen, the $f$ dependence of $\mu_0H_\text{R}$ is almost identical for the different samples, indicating the minor change of the magnetic properties of the Py layer due to the different bottom layers. In fact, by fitting the experimental data using Kittel formula~\cite{kittel1948theory}, 2$\pi$$f$/$\gamma=\sqrt{\mu_0 H_\mathrm{R}(\mu_0 H_\mathrm{R}+\mu_0 M_\mathrm{s})}$, the saturation magnetization is obtained to be $\mu_0 M_\text{s} = 0.746$, 0.753 and 0.777 T for the Py, Py/Pt(O) and Py/Cu/Pt(O) films, respectively. The minor difference ($< 5 \%$) in the saturation magnetization indicates that the larger damping of the Py/Pt(O) film cannot be attributed to possible different magnetic properties of the Py layer. Thus, the larger magnetic damping of the Py/Pt(O) film can only be attributed to the efficient absorption of the spin current at the interface. Notable is that the additional damping due to the spin-current absorption disappears by inserting the 3.6 nm-thick Cu layer between the Py and Pt(O) layers. Here, the thickness of the Cu layer is much thinner than its spin-diffusion length ($\sim 500$ nm)\cite{wang2014scaling}, allowing us to neglect the relaxation of the spin current in the Cu layer. This indicates that the direct contact between the Py and Pt(O) layers is essential for the absorption of the spin current at the interface, or the interface SOC.

\begin{figure}[tb]
\includegraphics[scale=1]{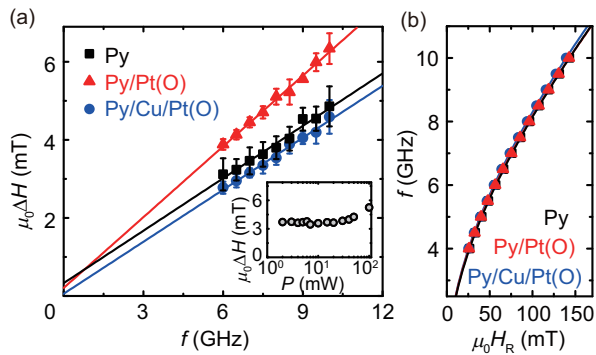}
\caption{
(a) The RF current frequency $f$ dependence of the half-width at half-maximum $\mu_0\Delta H$ for the Py, Py/Pt(O) and Py/Cu/Pt(O) samples. The solid lines are the linear fit to the experimental data. The inset shows RF current power $P$ dependence of $\mu_0\Delta H$ for the Py film at $f=7$ GHz. (b) The RF current frequency $f$ dependence of the resonance field $\mu_0H_\mathrm{R}$ for the three samples. The solid curves are the fitting result using the Kittel formula.  
}
\label{fig3} 
\end{figure}

\begin{table}[t]
  \caption{The summarized spin-absorption parameter $\Gamma_0\eta$ in different material systems. In order to directly compare this work with previous works, we used International System of Units. We used the magnetic permeability in vacuum $\mu_0= 4 \pi\times$10$^{-7}$ H/m. $\Delta \alpha$ and $\Gamma_0\eta$ for the Sn$_{0.02}$-Bi$_{1.08}$Sb$_{0.9}$Te$_2$S/Ni$_{81}$Fe$_{19}$ is the values at $T<100$ K.}
  \label{data}
  \centering
  \begin{tabular}{lccc}
    \hline \hline
    Heterostructure  & $\Delta\alpha$ & $\Gamma_0\eta$ [1/m$^2$]  & Ref. \\
    \hline
    Bi/Ag/Ni$_{80}$Fe$_{20}$  & 0.015  &  8.7$\times$10$^{18}$ & [\onlinecite{AgBi2016}]\\
    Bi$_2$O$_3$/Cu/Ni$_{80}$Fe$_{20}$  & 0.0045  &  1.5$\times$10$^{18}$ &  [\onlinecite{karube2016}]\\
    SrTiO$_{3}$/LaAlO$_{3}$/Ni$_{81}$Fe$_{19}$  & 0.0013  &  2.3$\times$10$^{18}$ & [\onlinecite{lesne2016highly}]\\
    Pt(O)/Ni$_{81}$Fe$_{19}$  & 0.0044  & 2.3$\times$10$^{18}$ & This work \\
    $\alpha$-Sn/Ag/Fe  & 0.022  &  1.2$\times$10$^{19}$ & [\onlinecite{rojas2016spin}]\\
    Sn$_{0.02}$-Bi$_{1.08}$Sb$_{0.9}$Te$_2$S/Ni$_{81}$Fe$_{19}$ & 0.013  & 1.4$\times$10$^{19}$ & [\onlinecite{nomura2017absorption}] \\
    Bi$_2$Se$_3$/Ni$_{81}$Fe$_{19}$  & 0.0013  &  2.5$\times$10$^{18}$ & [\onlinecite{deorani2014observation}]\\
    \hline \hline
  \end{tabular}
\end{table}

To quantitatively discuss the spin absorption at the Py/Pt(O) interface and compare with other material systems, we calculate the spin absorption parameters. In a model of the spin pumping where the interface SOC is taken into account, the additional damping constant is expressed as~\cite{chen2015spin}
\begin{equation}
\Delta \alpha= \frac{g\mu_\textrm{B}\Gamma_0}{\mu_0 M_\text{s} d}\left(\frac{1+6\eta\xi}{1+\xi}+\frac{\eta}{2(1+\xi)^2} \right). \label{alphacal}
\end{equation}
Here, $g=2.11$ is the $g$ factor\cite{shaw2013precise}, $\mu_\textrm{B}=9.27 \times 10 ^{-24}$ Am$^2$ is the Bohr magneton, $d$ is the thickness of the Py layer, and $\Gamma_0$ is the mixing conductance at the interface. $\xi$ is the back flow factor; no backflow refers to $\xi = 0$ and $\xi = \infty$ indicates that the entire spin current pumped into the bulk flows back across the interface. $\eta$ is the parameter that characterizes the interface SOC. For the Py/Pt(O) film, $\xi$ can be approximated to be $\infty$ because of the spin pumping into the bulk of the semi-insulating Pt(O) layer can be neglected. Thus, Eq.~(\ref{alphacal}) can be simplified as       
\begin{equation}
\Delta \alpha= \frac{6g\mu_\textrm{B}\Gamma_0\eta}{\mu_0 M_s d}.\label{LTa}
\end{equation}
Here, $6 \Gamma_0\eta$ corresponds to the effective spin mixing conductance $g_\text{eff}^{\uparrow\downarrow}$. From the enhancement of magnetic damping $\Delta \alpha$, we obtain $\Gamma_0\eta= 2.3 \times$ 10$^{18}$ m$^{-2}$ for the Py/Pt(O) film. We further compared this value with $\Gamma_0\eta$ for other systems where efficient interface charge-spin conversion has been reported. As shown in Table I, the spin-absorption parameter $\Gamma_0\eta$ of the Py/Pt(O) film is comparable with that of the heterostructures with the strong SOC, such as the 2DEG at an oxide interface, topological insulators, as well as metal/oxide and metallic junctions with the Rashba SOC. This result therefore demonstrates the strong SOC at the Py/Pt(O) interface.

\section{conclusions}
In summary, we have investigated the spin-current absorption and relaxation at the ferromagnetic-metal/Pt-oxide interface. By measuring the magnetic damping for the  Py, Py/Pt(O) and Py/Cu/Pt(O) structures, we show that the direct contact between Py and Pt(O) is essential for the absorption of the spin current, or the sizable interface SOC. Furthermore, we found that the strength of the spin-absorption parameter at the Py/Pt(O) interface is comparable to the value for intensively studied heterostructures with strong SOC, such as 2DEG at an oxide interface, topological insulators, metallic junction with Rashba SOC. The comparable value with these material systems illustrates the strong SOC at the ferromagnetic-metal/Pt-oxide interface. This indicates that the oxidation of heavy metals provides a novel approach for the development of the energy-efficient spintronics devices based the SOC.

\begin{acknowledgments}
This work was supported by JSPS KAKENHI Grant Numbers 26220604, 26103004, the Asahi Glass Foundation, JGC-S Scholarship Foundation, and Spintronics Research Network of Japan (Spin-RNJ). H.A. is JSPS International Research Fellow (No.~P17066) and acknowledges the support from the JSPS Fellowship (Grant No. 17F17066).
\end{acknowledgments}

%

\end{document}